\documentclass[aps,prl,twocolumn,a4paper,10pt]{revtex4-1} %%% PRL FORMAT  

\usepackage[T1]{fontenc}		%%% T1 Font Encoding
\usepackage[english]{babel}		%%% LaTeX for English
\usepackage[latin1]{inputenc}	%%% Accentuated Fonts
\usepackage{times}

\usepackage{amsmath}			%%% Mathematics AMS package 
\usepackage{amssymb}			%%% Mathematics AMS package 
\usepackage{bbm}				%%% Font for ensembes

\usepackage[colorlinks=true, a4paper=true, pdfstartview=FitV,linkcolor=blue, citecolor=blue, urlcolor=blue]{hyperref}
\usepackage{graphicx}			%%% Include graphics
\usepackage{graphics}			%%% Include graphics
\DeclareGraphicsExtensions{.pdf}
\usepackage{color}		
\newcommand{\CPtwo}{{GL}^{(3)}}
\newcommand{\Ztwo}{{Z}_2}

\newcommand{\mbf}{\mathbf}
\newcommand{\Figref}[1]{Fig.~\ref{#1}}
\newcommand{\Eqref}[1]{\eqref{#1}}
\newcommand{\ie}{{\it i.e.~}} 

\newcommand{\Equation}[2]{\begin{equation}\label{#1}#2\end{equation}}
\newcommand{\Align}[2]{\begin{align}\label{#1}#2\end{align}}
\renewcommand\Re{\mathrm{Re}}
\renewcommand\Im{\mathrm{Im}}

\newcommand{\oz}{{(0)}}

\begin{document}

%%%%%%%%%%%%%%%%%%%%%%%%%%%%%%%%%%%%%%%%%%%%%%%%%%%%%%%%%%%%%%%%%%%%%%%%%%%%%%%%%%%%%%%%%%
%%%%%%%%%%%%%%%%%%%%%%%%%%%%%%%%%%%%%%%%%%%%%%%%%%%%%%%%%%%%%%%%%%%%%%%%%%%%%%%%%%%%%%%%%%
%%%% Title informations and authors
\title{Topological solitons in three-band superconductors with broken time reversal symmetry}

\author{Julien~Garaud,~Johan~Carlstr\"om~and~Egor~Babaev}
%\email{garaud.phys@gmail.com}
\affiliation{ 
Department of Physics, University of Massachusetts Amherst, MA 01003 USA \\
Department of Theoretical Physics, The Royal Institute of Technology, Stockholm, SE-10691 Sweden
}
\date{\today}

%%%%%%%%%%%%%%%%%%%%%%%%%%%%%%%%%%%%%%%%%%%%%%%%%%%%%%%%%%%%%%%%%%%%%%%%%%%%%%%%%%%%%%%%%%
%%%% The abstract
\begin{abstract}
We show that three-band superconductors with broken time reversal symmetry allow magnetic flux-carrying stable 
topological solitons. They can be induced by 
fluctuations or quenching the system through a phase transition. It can provide an experimental signature of the time 
reversal symmetry breakdown.
\end{abstract}

\maketitle
%%%%%%%%%%%%%%%%%%%%%%%%%%%%%%%%%%%%%%%%%%%%%%%%%%%%%%%%%%%%%%%%%%%%%%%%%%%%%%%%%%%%%%%%%%
%%%%%%%%%%%%%%%%%%%%%%%%%%%%%%%%%%%%%%%%%%%%%%%%%%%%%%%%%%%%%%%%%%%%%%%%%%%%%%%%%%%%%%%%%%

Experiments on iron pnictide superconductors suggest the existence of more than two relevant superconducting 
bands \cite{iron2,*iron3,nagaosa,*stanev,*hu}. The new physics which can appear in these circumstances is the possible superconducting 
states with spontaneously broken time reversal symmetry (BTRS) as a consequence of frustration of competing 
interband Josephson couplings  \cite{nagaosa,*stanev,*hu} (other scenario for BTRS state was discussed in \cite{zhang,*platt}). 
BTRS states also attracted much interest earlier in the context of unconventional spin-triplet superconducting models. 
There  they have a different origin and are described by two-component Ginzburg-Landau models \cite{agterberg,*machida1,*machida2,*machida3}. 
In those cases the theory predicts  domain walls which pin vortices \cite{agterberg,*machida1,*machida2,*machida3}. It was suggested that this can 
result in formation of experimentally observable vortex sheets if (i) a domain wall itself is pinned by  sample inhomogeneities, 
or (ii) if a domain is dynamically formed inside a current-driven vortex lattice \cite{agterberg,*machida1,*machida2,*machida3}.

\begin{figure}[!htb]
\hbox to \linewidth{ \hss
   \includegraphics[width=0.33\linewidth]{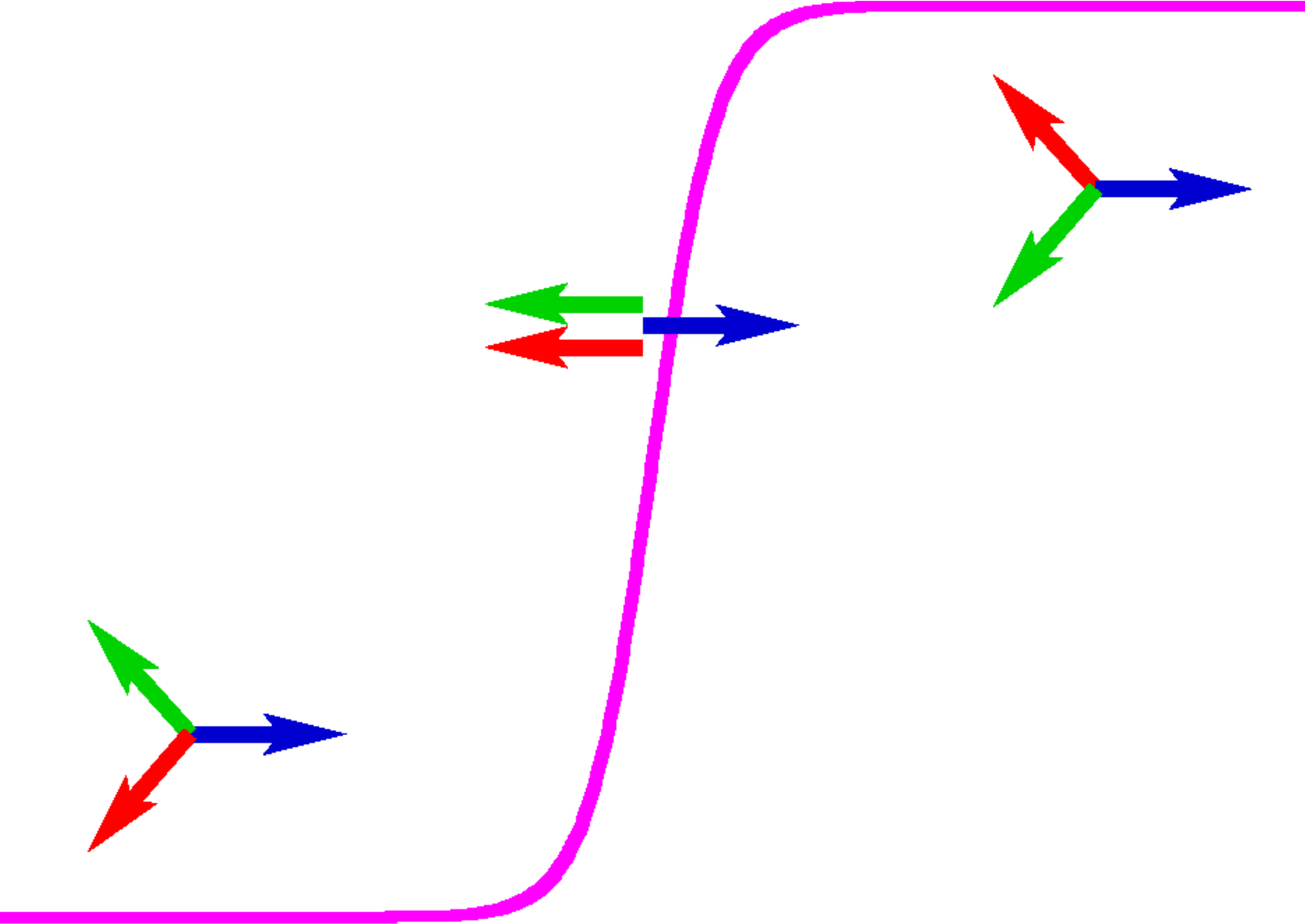}
   \includegraphics[width=0.33\linewidth]{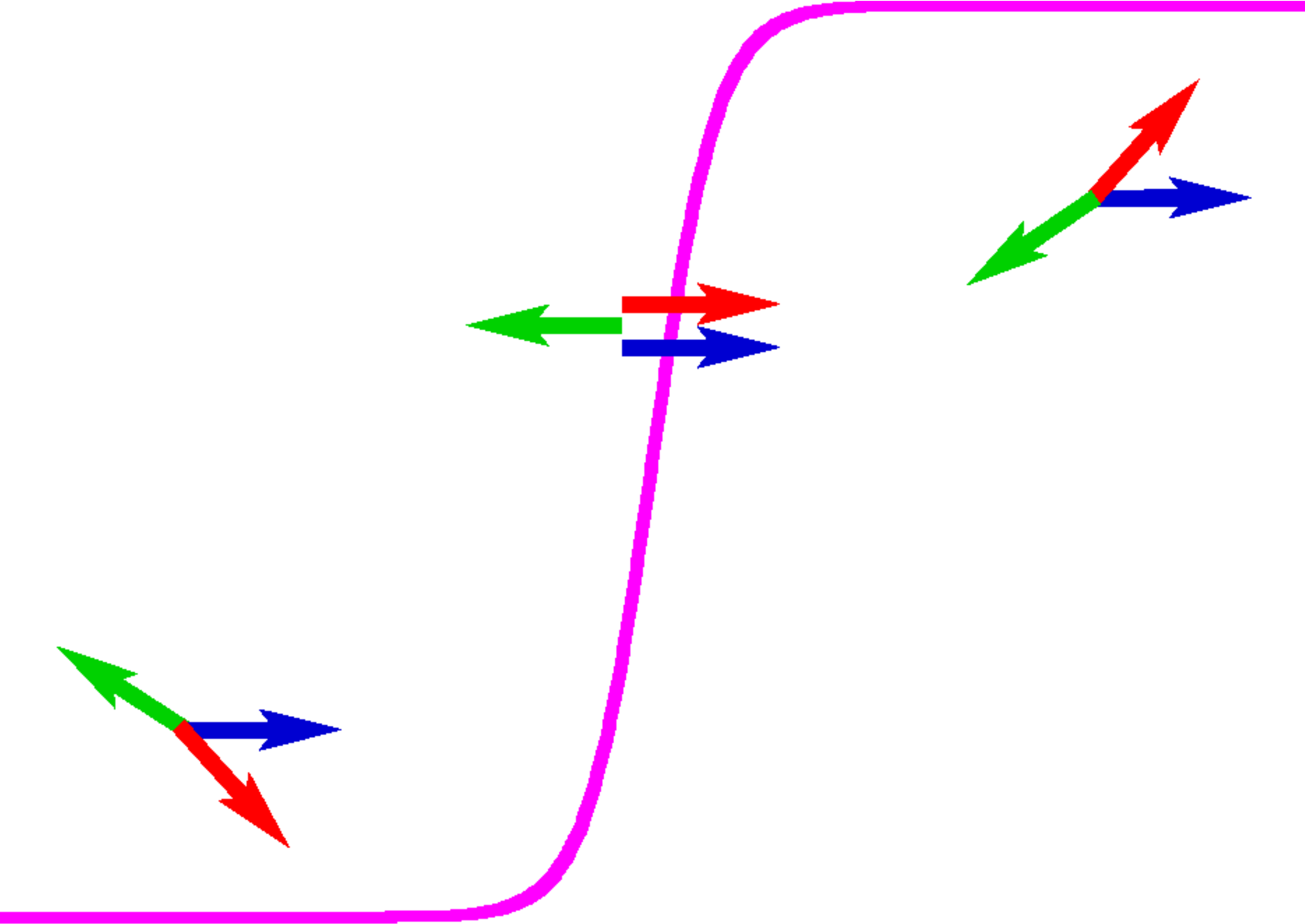}
   \includegraphics[width=0.33\linewidth]{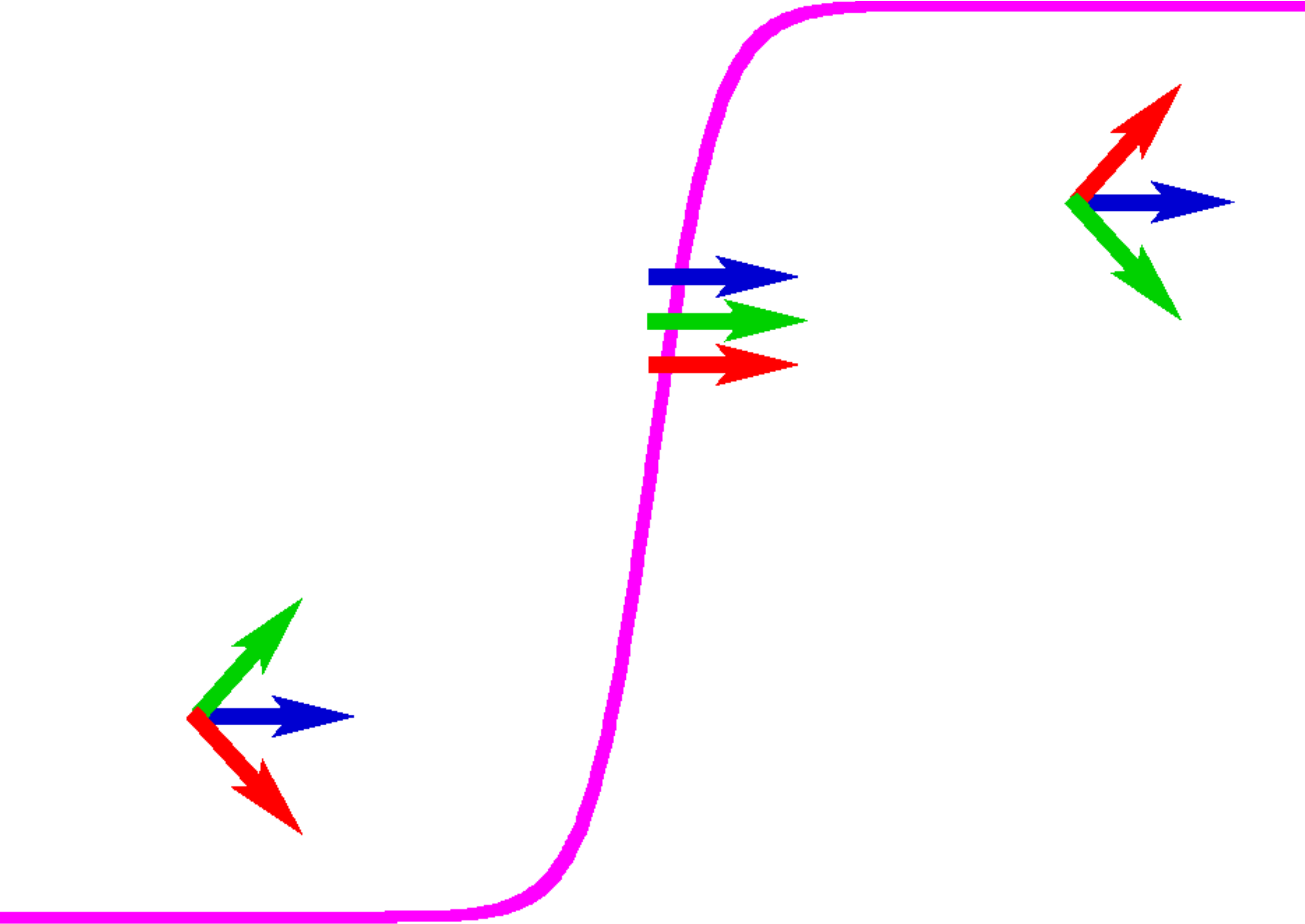}
\hss}
\caption{
(Color online) -- 
Schematic representation of various  $\Ztwo$ domain walls in three-band superconductors with different frustrations of 
phase angles, shown by arrows of different colors. Pink line schematically shows phase difference between red and 
green arrow, interpolating between the two inequivalent ground states.
}
\label{Domain}
\end{figure}

Here we show that a BTRS state in a three-band superconductor allows formation of metastable topological solitons. 
Although it is not by any means required to be near $T_c$ for these solitons to exist, we  use a {static} three-band 
Ginzburg-Landau (GL) free energy density model :
\begin{align} 
 F= &\frac{1}{2}(\nabla \times A)^2+ \sum_{i =1,2,3}\frac{1}{2}|D\psi_i|^2 
+V(\psi_i)	 \nonumber\\
-&\sum_{i =1,2,3}\sum_{j>i}\eta_{ij}|\psi_i|
|\psi_j|\cos(\varphi_i-\varphi_j)
\label{freeEnergy}
\end{align}
Here, $D=\nabla+ie{\bf A}$, and $\psi_i=|\psi_i| e^{i\varphi_i}$ are complex fields representing the superconducting 
components. We  choose to work here with a minimal  effective potential 
$V\equiv\sum_{i=1,2,3}\alpha_i|\psi_i|^2+\frac{1}{2}\beta_i|\psi_i|^4$. Although there could be various other terms 
allowed by symmetry in \Eqref{freeEnergy} they are not qualitatively important for  the discussion below. 
For $\eta_{ij}>0$, the Josephson interaction term is minimal for zero phase difference, while $\eta_{ij}<0$ it is minimal 
for $\varphi_i-\varphi_j=\pi$. When the signs of   $\eta_{ij}$ coefficients are all positive, [we denote it as $(+++)$] the 
ground state has $\varphi_1=\varphi_2=\varphi_3$. Similarly in case $(+--)$ one has phase locking pattern 
$\varphi_1=\varphi_2=\varphi_3+\pi$. However in cases  $(++-)$ and  $(---)$ there is a frustration between the phase 
locking tendencies [\ie one cannot simultaneously satisfy $\cos(\varphi_i-\varphi_j) = \pm 1$]. For example, consider  the 
case $\alpha_i=-1,\;\beta_i=1$ and $\eta_{ij}=-1$. Without loss of generality lets set $\varphi_1=0$ then two ground 
states are possible $\varphi_2=2\pi/3,\;\varphi_3=-2\pi/3$ or $\varphi_2=-2\pi/3,\;\varphi_3=2\pi/3$. Thus in these 
frustrated cases there is   $\Ztwo$ broken symmetry in the system associated with complex 
conjugation of the all $\psi$ fields. The broken $\Ztwo$ symmetry implies existence of domain walls solutions, which are 
schematically  shown on \Figref{Domain}.  Note that the frustrated phase differences can assume values different 
from $2\pi n/3$ in case of differing effective potentials or Josephson coupling 
strengths.

Let us  now outline basic properties of the model  \Eqref{freeEnergy}. Without intercomponent Josephson coupling 
and $\alpha_i<0$, its symmetry is $[U(1)]^3$. Then  it allows three kinds of fractional flux vortices with logarithmically 
diverging energy \cite{smiseth,*frac} characterized by a phase winding in (\ie integral over a phase gradient around a vortex) 
$\Delta \varphi_i \equiv \oint_\sigma \nabla \varphi_i = 2\pi$. Such a vortex carries a fraction of magnetic flux quanta 
($\Phi_0$), given by $\Phi_i = |\psi_i|^2/(|\psi_1|^2+|\psi_2|^2+|\psi_3|^2) \Phi_0$. However a bound state of three such 
vortices ($i=1,2,3$) has a finite energy. The finite-energy bound state is a ``composite" vortex which has one core singularity 
where $|\psi_1|+|\psi_2|+|\psi_3|=0$. Around this core all three phases have similar winding $\Delta \varphi_i=2\pi$. Thus it is 
a logarithmically bound state of fractional vortices whose flux adds up to one flux quantum $\Phi_0$. In case of non-zero 
Josephson coupling fractional vortices are bound much stronger since they interact linearly \cite{smiseth,*frac}. 

We show below that the model \Eqref{freeEnergy} remarkably  has a different kind of stable topological excitations 
distinct from vortices. Note that in two-component superconductors Skyrmion and Hopfion topological solitons can be 
represented as  bound states of  two spatially separated fractional vortices \cite{prb09}. Likewise we can represent a 
topological soliton {carrying $N$ flux quanta (\ie with each phase winding  $2\pi N$ )} in a three component 
superconductor like a stable bound state of {\it spatially separated } $3N$ fractional vortices. 
Below we will call it ``$\CPtwo$ soliton". At first glance, split fractional vortices could not be stable
in the model \Eqref{freeEnergy} because of the strong linear attractive interaction between fractional vortices caused by 
Josephson couplings. However we show that such solutions exist as topologically nontrivial {\it local} minima in the energy 
landscape of the model \Eqref{freeEnergy}. These solutions may also be viewed as combinations of fractional vortices 
and closed domain walls. 
 
Domain walls can form dynamically by a quench, but due to its line tension a single $Z_2$ closed domain wall 
(\ie a domain wall loop) should rapidly collapse. Because of the field gradients, the superfluid density is suppressed on a 
domain wall. Therefore it can pin vortices. Furthermore at a domain wall one has  energetically unfavorable values of 
cosines of phase differences  $\cos(\varphi_i-\varphi_j)$. Thus Josephson terms immediately at the domain wall energetically 
prefer to {\it split} integer flux vortices into fractional flux vortices since it allows to attain more favorable phase difference values
in between the split fractional vortices. (Note that, away from  domain walls,  Josephson terms give in contrast attractive 
interaction between fractional vortices). We find that if  the magnetic field penetration length is sufficiently large, then there is 
a length scale at which repulsion between the fractionalized vortices pinned by domain wall counterbalances the domain wall's 
tension. It thus results in a formation of a stable topological soliton made up of $3N$ fractional vortices. Thus these topological 
solitons represent a closed $Z_2$ domain wall along which there are  $N$  points of zeros of each condensate $|\psi_i|$. Around 
each of these zeros the phase $\varphi_i$ changes by $2\pi$. The total phase winding around the soliton is 
$\oint \nabla  \varphi_1 dl= \oint \nabla   \varphi_2dl =\oint \nabla   \varphi_3 dl=2\pi N$. Therefore it carries $N$ flux quanta.

\begin{figure}[!htb]
  \hbox to \linewidth{ \hss
  \includegraphics[width=\linewidth]{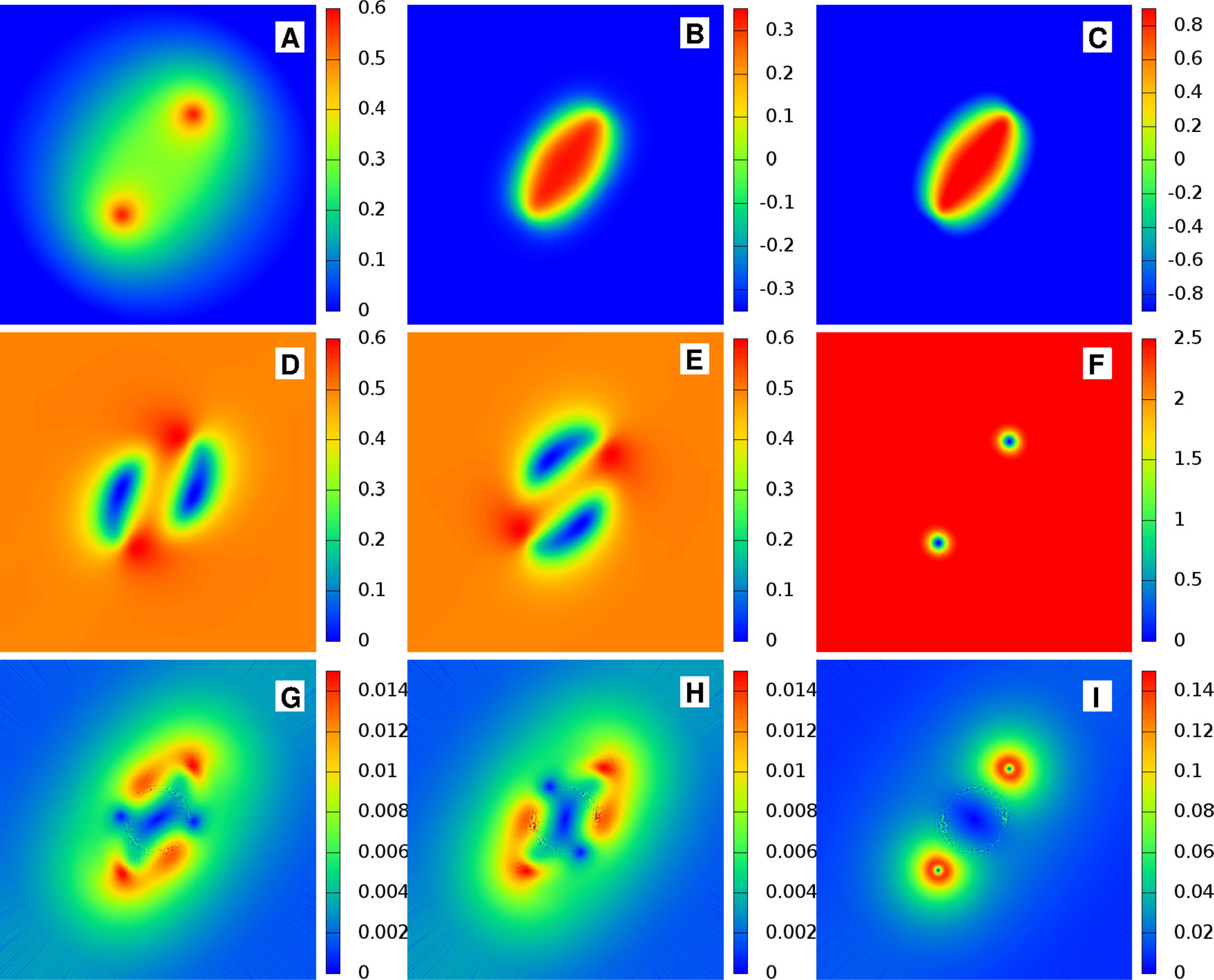}
 \hss}
\caption{
(Color online) -- 
$N=2$ topological solitons for two similar passive bands $(\alpha_i,\beta_i)=(1,1)$ with interband coupling $\eta_{12}=-3$. 
These bands have Josephson coupling $\eta_{13}=\eta_{23}=1$ to the third band, which is active $(\alpha_3,\beta_3)=(-2.5,1)$. 
The   system is   type-II  with $e=0.07$ (we use coupling constant $e$ in \Eqref{freeEnergy} to parametrize inverse penetration length). 
The panel $\mbf A$  displays the magnetic field ${B}$. Panels $\mbf B$  and $\mbf C$  respectively display $(\psi_1^*\psi_2-\psi_1\psi_2^*)/2i$ 
and $(\psi_1^*\psi_3-\psi_1\psi_3^*)/2i$, showing the phase difference between two condensates. Second line, shows the densities 
of the different condensates $|\psi_1|^2$ ($\mbf D$), $|\psi_2|^2$ ($\mbf E$), $|\psi_3|^2$ ($\mbf F$). The third line displays the 
supercurrent densities associated with each condensate $|J_1|$ ($\mbf G$), $|J_2|$ ($\mbf H$), $|J_3|$ ($\mbf I$). Phase differences 
on panels $\mbf B$ and $\mbf C$ show that there is a closed domain-wall since there are two areas with different phase-lockings 
(blue and red) associated with two possible ground states. The solution consists of $N=2$ vortices which are fractionalized : indeed,  
the panels $\mbf D$, $\mbf E$ and $\mbf F$ show separated highly asymmetric pairs of singularities  of different condensates. 
Note the very complicated geometry of  supercurrent densities shown  on  panels $\mbf G$, $\mbf H$ and $\mbf I$.
}
\label{Skyrmion-fig1}
\end{figure}

 Since it is a complicated nonlinear problem, no analytical tools are available and thus a conclusive answer if these solitons
 are stable could only be obtained numerically. We performed a numerical study based on energy minimization using a Non-Linear 
Conjugate Gradient algorithm showing the existence and stability of the $\CPtwo$ solitons. 
% Technical details of numerical calculations are available as supplementary online material. 
Technical details of numerical calculations are discussed in Appendix \ref{Numerics-fd}. 
The general tendency which we observed is, that in contrast to most of the known
topological solitons, they are more stable at higher topological charges. In fact we did not find any stable solitons for the lowest
topological charge corresponding to enclosed one quanta of magnetic flux ($N=1$). The  lowest topological charge solutions 
we found carry two flux quanta, and thus consist of six fractional vortices residing on a closed domain wall. The \Figref{Skyrmion-fig1} 
shows the $N=2$ soliton in a superconductor with two passive bands  (thus in this respect, similar to the models which are believed to be relevant 
for iron pnictide) coupled to an active band. Although it consists of six fractional vortices, one of the bands in this example has larger 
density and thus the magnetic field has two pronounced peaks near singularities in the main band. This is because the fractional
vortices in that band carry the largest amount of the magnetic flux $\Phi_3=|\psi_3|^2/[|\psi_1|^2+|\psi_2|^2+|\psi_3|^2]$. 
So the magnetic field profile of this soliton resembles a vortex pair. We similarly found $N=2$ solitons for superconductor 
with three passive bands and for three active bands which was not qualitatively different from the one shown on \Figref{Skyrmion-fig1}. 
 
\begin{figure}[!htb]
  \hbox to \linewidth{ \hss
  \includegraphics[width=\linewidth]{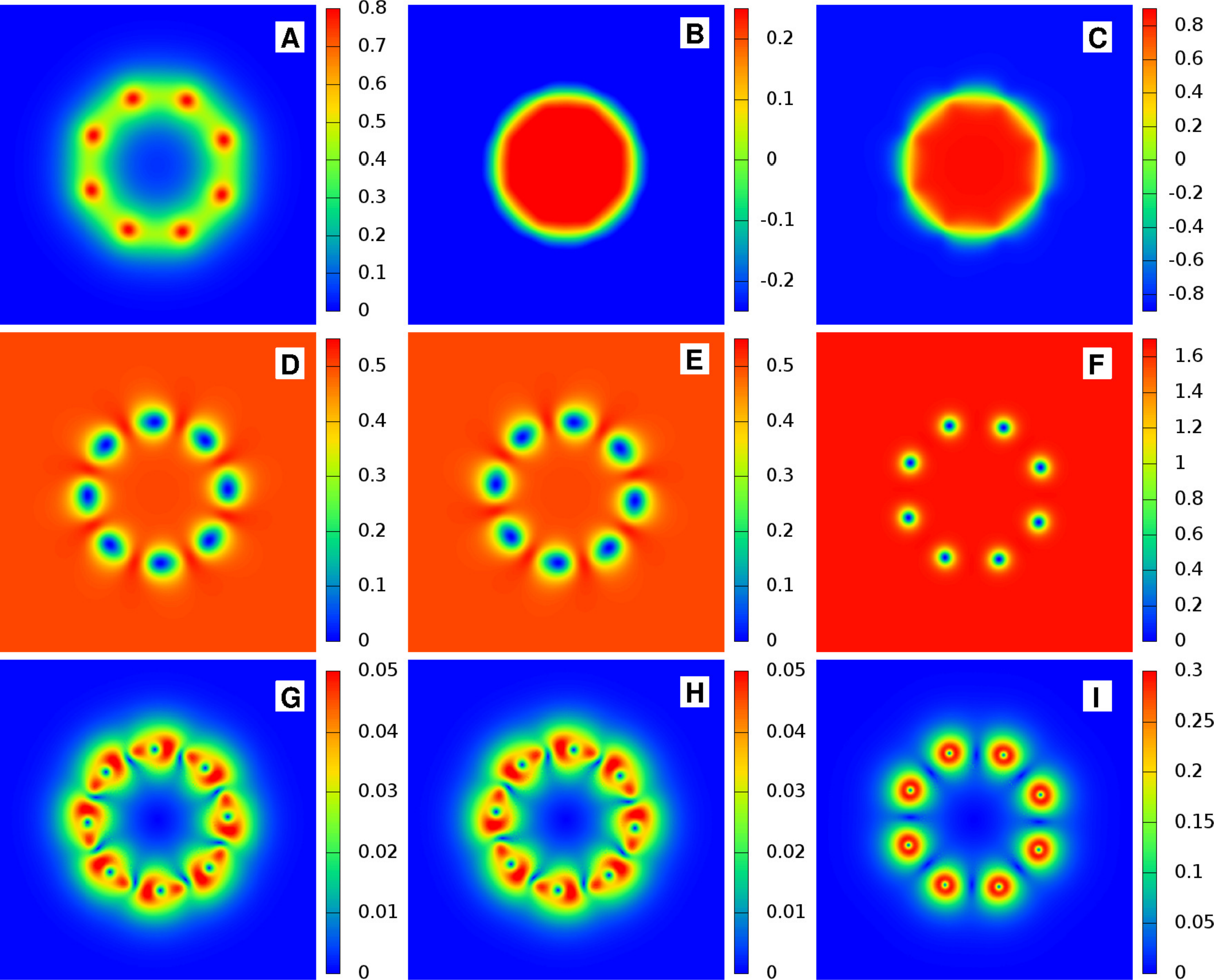}
  \hss}
\caption{%$e=0.3$
(Color online) -- 
$N=8$ quanta soliton for the same parameter set as in \Figref{Skyrmion-fig1} except that $e=0.3$ and $(\alpha_3,\beta_3)=(-1.5,1)$, 
giving less disparity in the ground state densities (displayed quantities are the same as in \Figref{Skyrmion-fig1}). The cores of vortices 
in each bands do not coincide. Note the complicated  structure of currents in each band.
}
\label{Skyrmion-fig2}
\end{figure}

We find that solutions with larger number of flux quanta tend to have  ring-like shapes. The \Figref{Skyrmion-fig2} gives an 
example of a solution with $N=8$ flux quanta. Note that this object will have a very distinct magnetic signature which can be 
distinguished by scanning SQUID or Hall or magnetic force microscopy. Despite the fact that this object is a bound state of 24 
fractional vortices, the magnetic field has only 8 pronounced maxima. They coincide with the position of the 8 singularities in 
the band with the largest density. 

\begin{figure}[!htb]
  \hbox to \linewidth{ \hss
  \includegraphics[width=\linewidth]{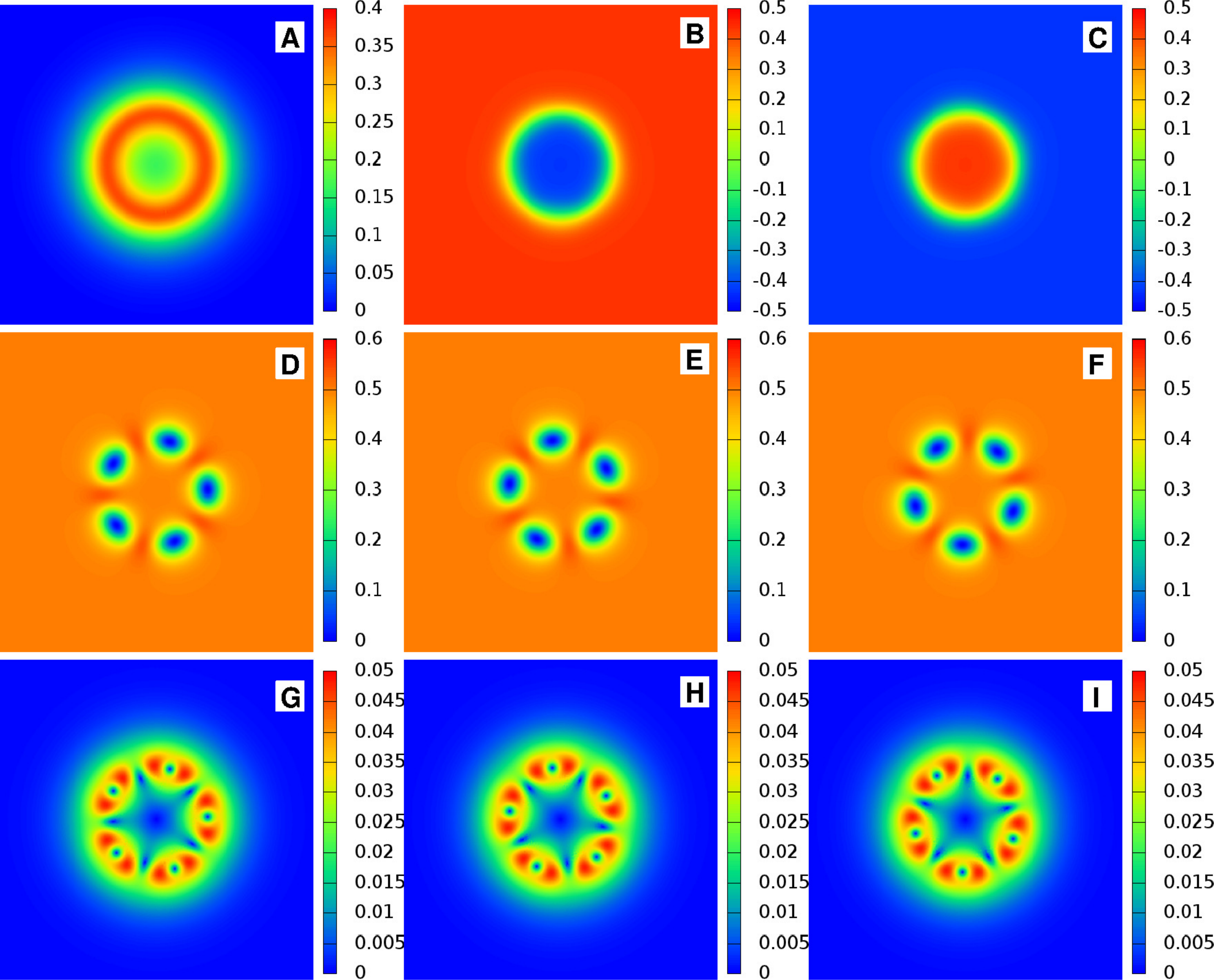}
  \hss}
\caption{
(Color online) -- 
$N=5$ quanta soliton with $e=0.3$. With three identical passive bands $(\alpha_i,\beta_i)=(1,1)$, with 
superconductivity induced by repulsion $\eta_{ij}=-3$ between the three condensates. Displayed quantities 
are the same as in \Figref{Skyrmion-fig1}.
}
\label{Skyrmion-fig4}
\end{figure}

The magnetic structure of the soliton always clearly reflects the relative densities the bands. When the ground state densities in 
each band are equal, the magnetic field has a uniform ring-like geometry as shown on \Figref{Skyrmion-fig4}. 

\begin{figure}[!htb]
  \hbox to \linewidth{ \hss
  \includegraphics[width=\linewidth]{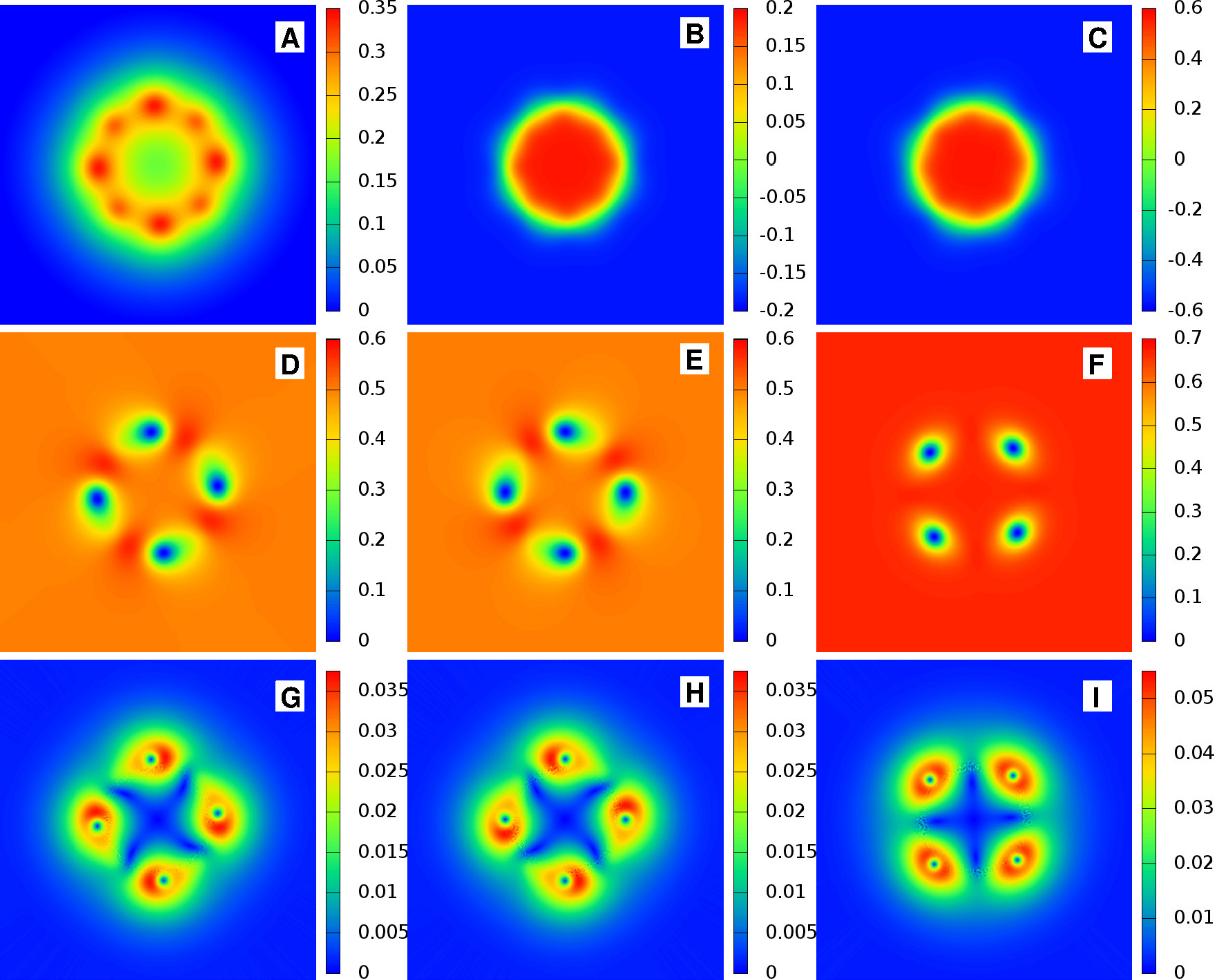}
  \hss}
\caption{
(Color online) -- 
$N=4$ quanta soliton  for two similar passive bands coupled to a third active band. The parameter set used 
here is the same as in \Figref{Skyrmion-fig2} except $(\alpha_3,\beta_3)=(-0.5,1)$ and $e=0.2$.
Displayed quantities are the same as in \Figref{Skyrmion-fig1}.
}
\label{Skyrmion-fig6}
\end{figure}

When disparity of the densities in different bands is small there is also a family of  $N$ quanta solitons which have 
$2N$ pronounced maxima in the magnetic field. An example with $N=4$ is shown on \Figref{Skyrmion-fig6}.

\begin{figure}[!htb]
\hbox to \linewidth{ \hss
%%%%%%%%%%%%%%%%%%%%%%%%%%%%%%%%%%%%%%%%%%
\begingroup
  \makeatletter
  \providecommand\color[2][]{%
    \GenericError{(gnuplot) \space\space\space\@spaces}{%
      Package color not loaded in conjunction with
      terminal option `colourtext'%
    }{See the gnuplot documentation for explanation.%
    }{Either use 'blacktext' in gnuplot or load the package
      color.sty in LaTeX.}%
    \renewcommand\color[2][]{}%
  }%
  \providecommand\includegraphics[2][]{%
    \GenericError{(gnuplot) \space\space\space\@spaces}{%
      Package graphicx or graphics not loaded%
    }{See the gnuplot documentation for explanation.%
    }{The gnuplot epslatex terminal needs graphicx.sty or graphics.sty.}%
    \renewcommand\includegraphics[2][]{}%
  }%
  \providecommand\rotatebox[2]{#2}%
  \@ifundefined{ifGPcolor}{%
    \newif\ifGPcolor
    \GPcolortrue
  }{}%
  \@ifundefined{ifGPblacktext}{%
    \newif\ifGPblacktext
    \GPblacktexttrue
  }{}%
  % define a \g@addto@macro without @ in the name:
  \let\gplgaddtomacro\g@addto@macro
  % define empty templates for all commands taking text:
  \gdef\gplbacktext{}%
  \gdef\gplfronttext{}%
  \makeatother
  \ifGPblacktext
    % no textcolor at all
    \def\colorrgb#1{}%
    \def\colorgray#1{}%
  \else
    % gray or color?
    \ifGPcolor
      \def\colorrgb#1{\color[rgb]{#1}}%
      \def\colorgray#1{\color[gray]{#1}}%
      \expandafter\def\csname LTw\endcsname{\color{white}}%
      \expandafter\def\csname LTb\endcsname{\color{black}}%
      \expandafter\def\csname LTa\endcsname{\color{black}}%
      \expandafter\def\csname LT0\endcsname{\color[rgb]{1,0,0}}%
      \expandafter\def\csname LT1\endcsname{\color[rgb]{0,1,0}}%
      \expandafter\def\csname LT2\endcsname{\color[rgb]{0,0,1}}%
      \expandafter\def\csname LT3\endcsname{\color[rgb]{1,0,1}}%
      \expandafter\def\csname LT4\endcsname{\color[rgb]{0,1,1}}%
      \expandafter\def\csname LT5\endcsname{\color[rgb]{1,1,0}}%
      \expandafter\def\csname LT6\endcsname{\color[rgb]{0,0,0}}%
      \expandafter\def\csname LT7\endcsname{\color[rgb]{1,0.3,0}}%
      \expandafter\def\csname LT8\endcsname{\color[rgb]{0.5,0.5,0.5}}%
    \else
      % gray
      \def\colorrgb#1{\color{black}}%
      \def\colorgray#1{\color[gray]{#1}}%
      \expandafter\def\csname LTw\endcsname{\color{white}}%
      \expandafter\def\csname LTb\endcsname{\color{black}}%
      \expandafter\def\csname LTa\endcsname{\color{black}}%
      \expandafter\def\csname LT0\endcsname{\color{black}}%
      \expandafter\def\csname LT1\endcsname{\color{black}}%
      \expandafter\def\csname LT2\endcsname{\color{black}}%
      \expandafter\def\csname LT3\endcsname{\color{black}}%
      \expandafter\def\csname LT4\endcsname{\color{black}}%
      \expandafter\def\csname LT5\endcsname{\color{black}}%
      \expandafter\def\csname LT6\endcsname{\color{black}}%
      \expandafter\def\csname LT7\endcsname{\color{black}}%
      \expandafter\def\csname LT8\endcsname{\color{black}}%
    \fi
  \fi
  \setlength{\unitlength}{0.0500bp}%
\resizebox{115pt}{84pt}{   \begin{picture}(7200.00,5040.00)%
    \gplgaddtomacro\gplbacktext{%
      \csname LTb\endcsname%
      \put(1078,704){\makebox(0,0)[r]{\strut{}\huge 1}}%
%      \put(1078,1213){\makebox(0,0)[r]{\strut{}\huge 1.02}}%
      \put(1078,1722){\makebox(0,0)[r]{\strut{}\huge 1.04}}%
%      \put(1078,2231){\makebox(0,0)[r]{\strut{}\huge 1.06}}%
      \put(1078,2740){\makebox(0,0)[r]{\strut{}\huge 1.08}}%
%      \put(1078,3248){\makebox(0,0)[r]{\strut{}\huge 1.1}}%
      \put(1078,3757){\makebox(0,0)[r]{\strut{}\huge 1.12}}%
%      \put(1078,4266){\makebox(0,0)[r]{\strut{}\huge 1.14}}%
      \put(1078,4775){\makebox(0,0)[r]{\strut{}\huge 1.16}}%
      \put(1210,484){\makebox(0,0){\strut{}\huge 1}}%
      \put(2009,484){\makebox(0,0){\strut{}\huge 2}}%
      \put(2808,484){\makebox(0,0){\strut{}\huge 3}}%
      \put(3607,484){\makebox(0,0){\strut{}\huge 4}}%
      \put(4406,484){\makebox(0,0){\strut{}\huge 5}}%
      \put(5205,484){\makebox(0,0){\strut{}\huge 6}}%
      \put(6004,484){\makebox(0,0){\strut{}\huge 7}}%
      \put(6803,484){\makebox(0,0){\strut{}\huge 8}}%
      \put(176,2739){\rotatebox{-270}{\makebox(0,0){\strut{}\huge $E$}}}%
      \put(4006,154){\makebox(0,0){\strut{}\huge $N$}}%
  
     \put(4200,1100){\makebox(0,0){\strut{}\huge $e=1.1$}}%
     \put(2800,1900){\makebox(0,0){\strut{}\huge $e=0.9$}}%
     \put(2000,2700){\makebox(0,0){\strut{}\huge $e=0.7$}}%
     \put(2000,3400){\makebox(0,0){\strut{}\huge $e=0.5$}}%
     \put(2000,4084){\makebox(0,0){\strut{}\huge $e=0.3$}}%
  }%
    \gplgaddtomacro\gplfronttext{%
    }%
    \gplbacktext
    \put(0,0){\includegraphics{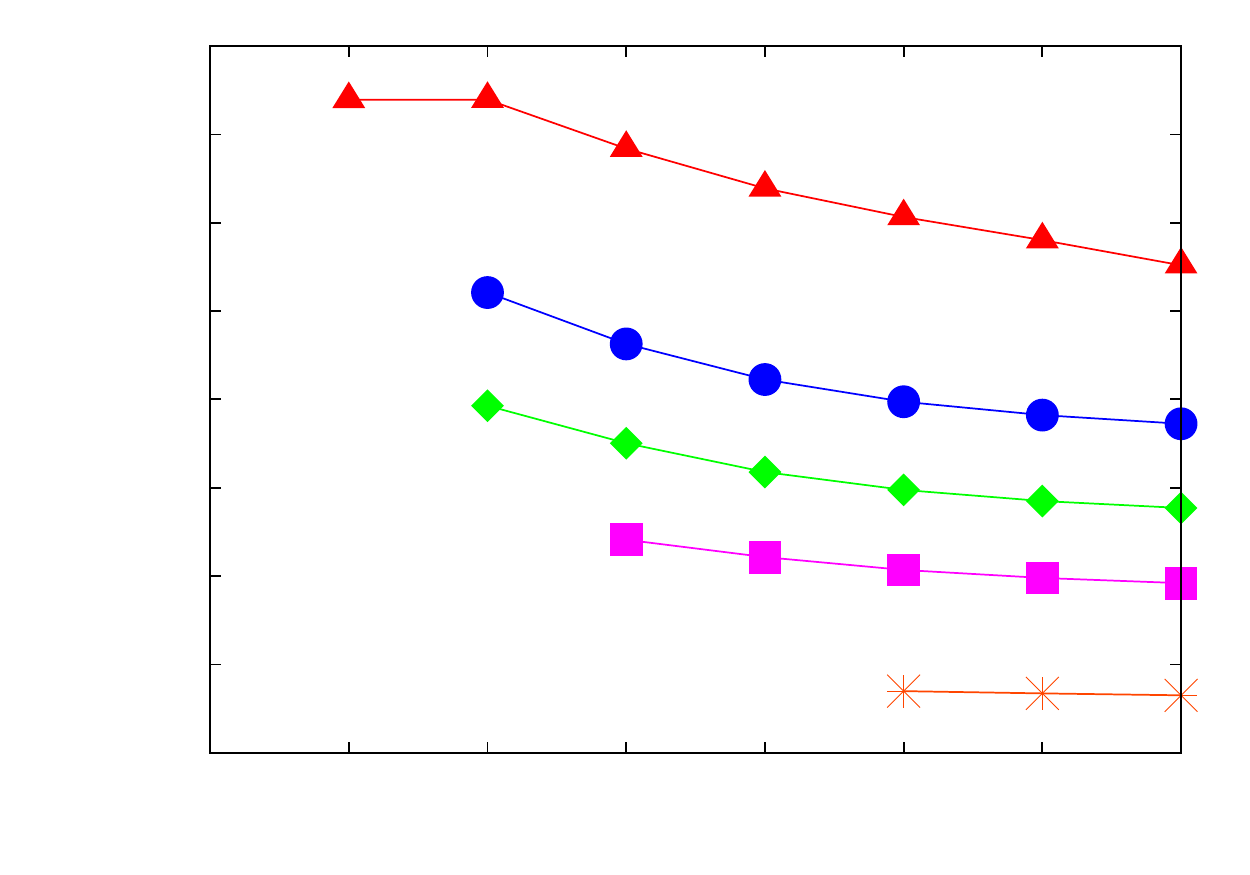}}%
    \gplfronttext
  \end{picture}}%
\endgroup
%%%%%%%%%%%%%%%%%%%%%%%%%%%%%%%%%%%%%%%%%%
% GNUPLOT: LaTeX picture with Postscript
\begingroup
  \makeatletter
  \providecommand\color[2][]{%
    \GenericError{(gnuplot) \space\space\space\@spaces}{%
      Package color not loaded in conjunction with
      terminal option `colourtext'%
    }{See the gnuplot documentation for explanation.%
    }{Either use 'blacktext' in gnuplot or load the package
      color.sty in LaTeX.}%
    \renewcommand\color[2][]{}%
  }%
  \providecommand\includegraphics[2][]{%
    \GenericError{(gnuplot) \space\space\space\@spaces}{%
      Package graphicx or graphics not loaded%
    }{See the gnuplot documentation for explanation.%
    }{The gnuplot epslatex terminal needs graphicx.sty or graphics.sty.}%
    \renewcommand\includegraphics[2][]{}%
  }%
  \providecommand\rotatebox[2]{#2}%
  \@ifundefined{ifGPcolor}{%
    \newif\ifGPcolor
    \GPcolortrue
  }{}%
  \@ifundefined{ifGPblacktext}{%
    \newif\ifGPblacktext
    \GPblacktexttrue
  }{}%
  % define a \g@addto@macro without @ in the name:
  \let\gplgaddtomacro\g@addto@macro
  % define empty templates for all commands taking text:
  \gdef\gplbacktext{}%
  \gdef\gplfronttext{}%
  \makeatother
  \ifGPblacktext
    % no textcolor at all
    \def\colorrgb#1{}%
    \def\colorgray#1{}%
  \else
    % gray or color?
    \ifGPcolor
      \def\colorrgb#1{\color[rgb]{#1}}%
      \def\colorgray#1{\color[gray]{#1}}%
      \expandafter\def\csname LTw\endcsname{\color{white}}%
      \expandafter\def\csname LTb\endcsname{\color{black}}%
      \expandafter\def\csname LTa\endcsname{\color{black}}%
      \expandafter\def\csname LT0\endcsname{\color[rgb]{1,0,0}}%
      \expandafter\def\csname LT1\endcsname{\color[rgb]{0,1,0}}%
      \expandafter\def\csname LT2\endcsname{\color[rgb]{0,0,1}}%
      \expandafter\def\csname LT3\endcsname{\color[rgb]{1,0,1}}%
      \expandafter\def\csname LT4\endcsname{\color[rgb]{0,1,1}}%
      \expandafter\def\csname LT5\endcsname{\color[rgb]{1,1,0}}%
      \expandafter\def\csname LT6\endcsname{\color[rgb]{0,0,0}}%
      \expandafter\def\csname LT7\endcsname{\color[rgb]{1,0.3,0}}%
      \expandafter\def\csname LT8\endcsname{\color[rgb]{0.5,0.5,0.5}}%
    \else
      % gray
      \def\colorrgb#1{\color{black}}%
      \def\colorgray#1{\color[gray]{#1}}%
      \expandafter\def\csname LTw\endcsname{\color{white}}%
      \expandafter\def\csname LTb\endcsname{\color{black}}%
      \expandafter\def\csname LTa\endcsname{\color{black}}%
      \expandafter\def\csname LT0\endcsname{\color{black}}%
      \expandafter\def\csname LT1\endcsname{\color{black}}%
      \expandafter\def\csname LT2\endcsname{\color{black}}%
      \expandafter\def\csname LT3\endcsname{\color{black}}%
      \expandafter\def\csname LT4\endcsname{\color{black}}%
      \expandafter\def\csname LT5\endcsname{\color{black}}%
      \expandafter\def\csname LT6\endcsname{\color{black}}%
      \expandafter\def\csname LT7\endcsname{\color{black}}%
      \expandafter\def\csname LT8\endcsname{\color{black}}%
    \fi
  \fi
  \setlength{\unitlength}{0.0500bp}%
 \resizebox{115pt}{84pt}{ \begin{picture}(7200.00,5040.00)%
    \gplgaddtomacro\gplbacktext{%
      \csname LTb\endcsname%
      \put(946,804){\makebox(0,0)[r]{\strut{} \huge0}}%
  %    \put(946,1383){\makebox(0,0)[r]{\strut{}\huge 0.2}}%
      \put(946,2061){\makebox(0,0)[r]{\strut{} \huge0.4}}%
%      \put(946,2740){\makebox(0,0)[r]{\strut{}\huge 0.6}}%
      \put(946,3418){\makebox(0,0)[r]{\strut{}\huge 0.8}}%
  %    \put(946,4097){\makebox(0,0)[r]{\strut{}\huge 1}}%
     \put(946,4775){\makebox(0,0)[r]{\strut{}\huge 1.2}}%
      \put(1078,484){\makebox(0,0){\strut{}\huge -20}}%
 %     \put(1794,484){\makebox(0,0){\strut{}\huge-15}}%
      \put(2509,484){\makebox(0,0){\strut{}\huge-10}}%
      \put(3225,484){\makebox(0,0){\strut{}\huge-5}}%
      \put(3941,484){\makebox(0,0){\strut{}\huge 0}}%
      \put(4656,484){\makebox(0,0){\strut{}\huge 5}}%
      \put(5372,484){\makebox(0,0){\strut{}\huge 10}}%
 %     \put(6087,484){\makebox(0,0){\strut{}\huge 15}}%
      \put(6803,484){\makebox(0,0){\strut{}\huge 20}}%
      \put(176,2739){\rotatebox{-270}{\makebox(0,0){\strut{}\huge $B$}}}%
      \put(3940,154){\makebox(0,0){\strut{}\huge $x$}}%
    }%
    \gplgaddtomacro\gplfronttext{%
    }%
    \gplbacktext
    \put(0,0){\includegraphics{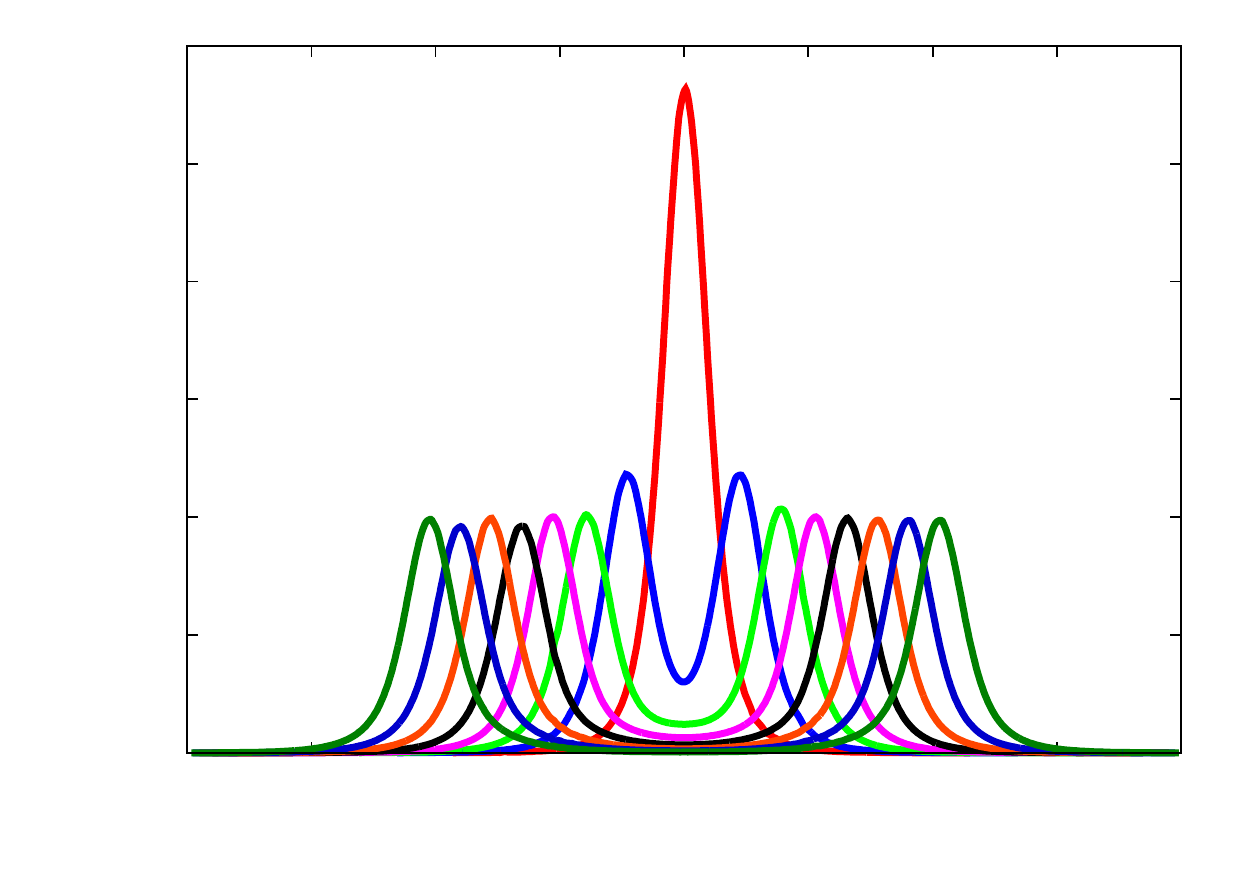}}%
    \gplfronttext
  \end{picture}}%
\endgroup
\hss}
\caption{
(Color online) --
Energies of the solitons per flux quanta, in the units of the energy of a single ordinary vortex (left). When the electric 
charge increases (\ie the penetration length decreases) solitons with smaller $N$ become unstable. The right panel 
shows crossections of the magnetic field for  solitons with $N\in[2,8]$ (double-peak curves). The central curve 
corresponds to a crossection of a regular $N=1$  vortex. The parameters of the Ginzburg-Landau model used here 
are the same as in \Figref{Skyrmion-fig4}, which gives nearly axially-symmetric magnetic field.
}
\label{Skyrmion-energy}
\end{figure}

We investigated numerically more than 500 parameter sets  in three-component BTRS GL models. For all type-II 
three-component BTRS GL models we found stable $\CPtwo$ solitons, provided the topological charge was large enough.
The solution existed  in BTRS states irrespectively of whether bands are active or passive and for very different 
effective potentials and interband coupling strengths. It indicates that these solitons should be rather generic 
excitations in three-component type-II BTRS superconductors. \Figref{Skyrmion-energy} shows the energy and 
stability of the solitons for different values of the coupling constant $e$ (in our parametrization $e$ controls the 
inverse magnetic field penetration length). It reflects the generic tendency which we find, that the solitons are more 
stable in more type-II regimes and also at higher topological charges. 

%%%%%%%%%%%%%%%%%%%%%%%%%%%%%%%%%%%%%%%%%%%%%%%%%%%%%%%%%%%%%%%%%%%%%%%%%%%%%%%%%%%%%%%%%%
%%%%%%%%%%%%%%%%%%%%%%%%%%%%%%%%%%%%%%%%%%%%%%%%%%%%%%%%%%%%%%%%%%%%%%%%%%%%%%%%%%%%%%%%%%

Lets us now address the physical observability of these solitons. First in all the cases which we studied in the model 
\Eqref{freeEnergy}, the solitons with $N$ flux quanta were more energetically expensive than $N$ isolated one-quanta vortices. 
However they are protected by an energy barrier against decay into ordinary vortices. 
Note that because   the solitons are obtained as solutions of the energy minimization problem, they are guaranteed to be stable 
against infinitesimally small 
perturbations. However, since they are more energetic than vortices, 
\emph{strong enough} perturbation  should destabilize  them. 
This stability question is addressed numerically  in the Appendix \ref{Numerics-stab}.
%Supplementary material. 
For strongly type-II regime the potential 
barrier can be estimated as the energy needed to disconnect the domain wall. For a soliton in a three-dimensional sample with phase
winding in the $xy$-plane the potential barrier can be estimated as [coherence length]${}^2\times$[sample size in the 
direction of applied magnetic field]$\times$[condensation energy density].

\begin{figure}[!htb]
  \hbox to \linewidth{ \hss
  \includegraphics[width=\linewidth]{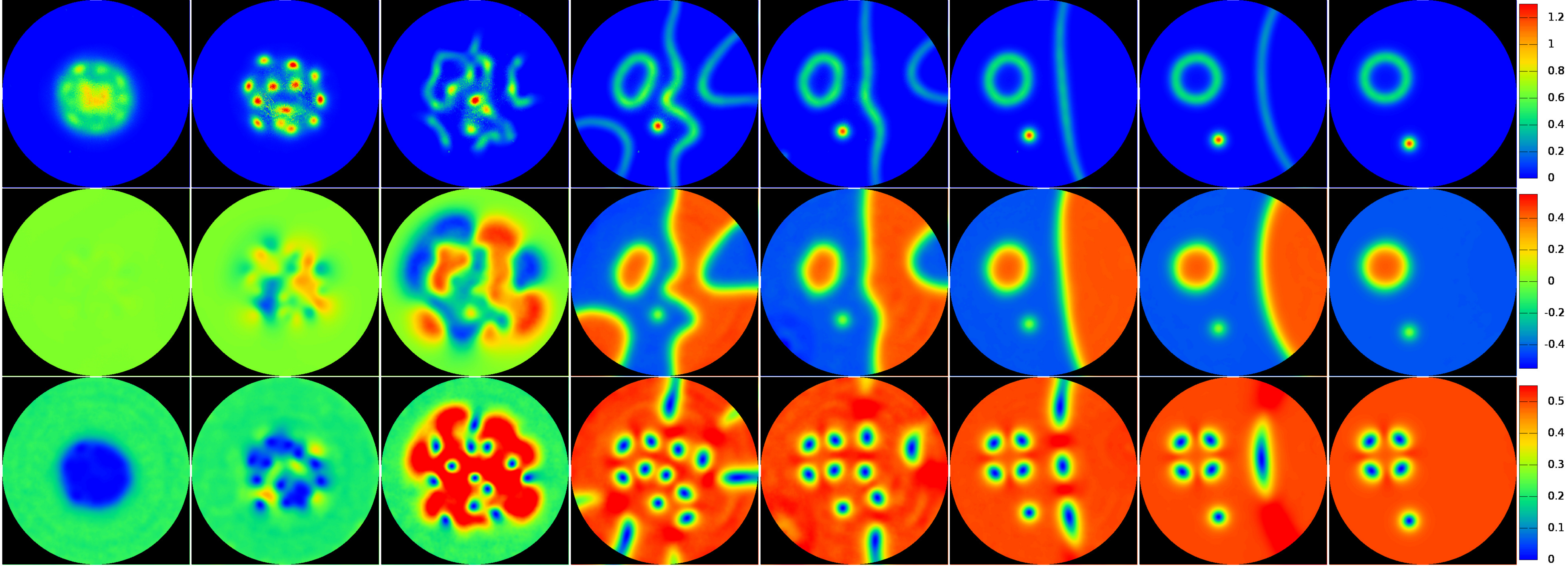}
  \hss}
\caption{
(Color online) -- 
The soliton formation during energy relaxation of an initial state of expanding group of vortices in a circular system 
with open boundary conditions. First line displays the  energy density. Second line shows the phase difference between 
condensates $(\psi_1^*\psi_2-\psi_1\psi_2^*)/2i$. When domain walls form they separate two inequivalent ground 
states (blue and red). Third line is the density of the first condensate $|\psi_1|^2$. Initial configuration has a high 
density of $13$ vortices in the center. Repulsive type-II interaction makes all vortices move away from each other
and escape the sample. In the process of energy minimization   domain walls and $\CPtwo$ solitons form. Domain 
wall connected to boundaries quickly disappear. The final picture shows the resulting long-living state of a well 
separated $N=4$ $\CPtwo$ soliton and a vortex. Parameter set used here is the same as in \Figref{Skyrmion-fig4}, 
with $e=0.4$.
}
\label{Formation-fig1}
\end{figure}

Being more expensive than vortices, these objects cannot  form as a 
ground state in low external field \footnote{ In principle by adding certain mixed gradient  terms, or density-density interaction 
 which gives energy penalty to the vortex cores where  the total density is zero
 (\ie  $\Sigma_{i} |\psi_i({\bf r})|^2$),  yields models where soliton lattice should 
 form instead of vortex lattice as a ground state in external field.}. 
However as  demonstrated in \Figref{Skyrmion-energy} they are not much 
more energetically expensive than vortices. In fact the corresponding energy differences can be just a few percent. Thus 
they can be excited by either by (a) thermal fluctuations or (b) by quenching in a sample subjected to a magnetic field. To 
address the  scenario (b) of possible formation of these solitons in a post-quench relaxation, we have to assess ``capture basin" 
of these solutions (\ie how large is the area in the free energy landscape from which an excited system would relax into the 
local minimum corresponding to a soliton. Although studying real post-quench relaxation dynamics is beyond the scope of 
this paper, nonetheless we can directly assess the capture basin of the solutions from the evolution of the system in our 
relaxation scheme (see also remark 
\footnote{We do not study dynamics in this paper. However we note that because Time Dependent 
Ginzburg-Landau (TDGL) equations can be seen as the gauge invariant gradient flow of the 
free energy, our numerical relaxation  scheme in fact is indirectly related to the TDGL 
dynamics of the system, see e.g. 
Q. Du, 
   {\it Journ. of  Math. Phys. \/}{\bf 46} 095109 (2005).}). We investigated several hundreds regimes and found that solitons typically 
easily form when a system is relaxed from various higher energy states. This indicates that the capture basin of these 
solutions is typically very large. We find that these defects in fact very easily form during  a rapid expansion of vortex 
lattice (which should occur  when magnetic field is rapidly lowered, or if a system is quenched through $H_{c2}$). A typical 
example is shown on \Figref{Formation-fig1}. Animations of these processes are available as a supplementary online material \cite{Julien}. 

\onecolumngrid

\twocolumngrid

In conclusion,  we have shown that BTRS state of a three-band superconductor can be detected through its magnetic 
response. Namely we have demonstrated that in this state the system has two kinds of flux carrying topological defects : 
ordinary vortices and also a different kind of topological solitons.  These solitons are only slightly more energetically 
expensive than vortices (in  some cases we found the energy difference as small as $10^{-2} E_v$ where $E_v$ is the 
energy of a vortex). They should form during a post-quench relaxation of a BTRS superconductor in an external field, since 
they represent local minima with a wide capture basin in the free energy landscape. I.e. a system should relax to these 
local minima from a wide variety of excited states. Then these solitons can be observed  in scanning SQUID, Hall, or magnetic 
force microscopy measurements. They can provide an experimental signature of  possible BTRS states in iron pnictide 
superconductors. A tendency for vortex  pair formation, yielding magnetic profile similar to that shown on \Figref{Skyrmion-fig1} 
was observed in $\rm Ba(Fe_{1-x}Co_x)_2As_2$, \cite{Kaliski} as well as vortex clustering in $\rm BaFe_{2-x}Ni_xAs_2$ \cite{Li}. 
These materials have strong pinning which can naturally produce disordered vortex states \cite{Li}, although a possibility of ``type-1.5" 
scenario for these vortex inhomogeneities was also voiced in \cite{Li}. The vortex pairs observed in \cite{Kaliski} can be discriminated 
from $N=2$ solitons (such as that shown on Fig. \ref{Skyrmion-fig1}), by quenching the system and observing whether or not it forms 
vortex triangles, squares, pentagons etc corresponding to higher-$N$ solitons.

We thank J.M. Speight for useful communications. The work is supported by the Swedish Research Council, and by the Knut and Alice Wallenberg
Foundation through the Royal Swedish Academy of Sciences fellowship and by NSF CAREER Award No. DMR-0955902. 

%%%%%%%%%%%%%%%%%%%%%%%%%%%%%%%%%%%%%%%%%%%%%%%%%%%%%%%%%%%%%%%%%%%%%%%%%%%%%%%%%%%%%%%%%%
%%%% Appendix
\appendix

\section{Appendix A : Finite element energy minimization}\label{Numerics-fd}

The $\CPtwo$ solitons are {\it local}  minima of the Ginzburg-Landau energy \Eqref{freeEnergy}. This means that 
functional minimization of \Eqref{freeEnergy}, from an appropriate initial guess carrying several flux quanta, should lead 
to a $\CPtwo$ soliton (if it exists as a stable solution). We consider the two-dimensional problem  \Eqref{freeEnergy} 
defined on the bounded domain $\Omega\subset\mathbbm{R}^2$, supplemented by a `open' boundary 
conditions on $\partial\Omega$.

Strictly speaking, there is a constraint on $\partial\Omega$. This `open constraint' is a particular Neumann boundary 
condition, such that the normal derivative of the fields on the boundary are zero. These boundary conditions in fact 
are a very weak constraint. For this problem one could also apply Robin boundary conditions on $\partial\Omega$, so 
that the fields satisfy linear asymptotic behavior (exponential localization). However, we choose to apply the `open' 
boundary conditions which are less constraining for the problem in question. `Open' boundary conditions also imply 
that topological defects can easily escape from the numerical grid, since it would further minimize the energy. To 
prevent this, the numerical grid is chosen to be large enough so that the attractive interaction with the boundaries 
is negligible. The size of the domain is then much larger than the typical interaction length scales. Thus in this method 
one has to use large numerical grids, which is computationally demanding. At the same time the advantage 
is that it is guaranteed that obtained solutions are not boundary pressure artifacts. 

The variational problem is defined for numerical computation using a finite element formulation provided by the 
Freefem++ library \cite{Hecht.Pironneau.ea}. Discretization within finite element formulation is done via a 
(homogeneous) triangulation over $\Omega$, based on Delaunay-Voronoi algorithm. Functions are decomposed on 
a continuous piecewise quadratic basis on each triangle. The accuracy of such method is controlled through the number 
of triangles, (we  typically used   $3\sim6\times10^4$), the order of expansion of the basis on each triangle (P2 elements 
being 2nd order polynomial basis on each triangle), and also the order of the quadrature formula for the integral on the 
triangles. 

Once the problem is mathematically well defined, a numerical optimization algorithm is used to solve the variational 
nonlinear problem (\ie to find the minima of $\mathcal{F}$). We used here a  Nonlinear Conjugate Gradient method. 
The algorithm is iterated until relative variation of the norm of the gradient of the functional  $\mathcal{F}$ with 
respect to all degrees of freedom is less than $10^{-6}$. 

\subsection{Initial guess}

As discussed in the paper, $N$ quanta $\CPtwo$ solitons in the three-component model are more energetically 
expensive than $N$ quanta ordinary vortices. They are \emph{local} minima of the energy functional \Eqref{freeEnergy}. 
As a result the initial guess should be within the attractive basin of the $\CPtwo$ solitons. Otherwise the configuration 
converges to ordinary vortices which have the same total phase winding but cost less energy. We find however the 
attractive basin of the $\CPtwo$  soliton solutions to be generally quite large (\ie the $\CPtwo$ soliton forms quite easily in general).
The initial field configuration carrying $N$ flux quanta is prepared by using an ansatz which imposes phase windings 
around spatially separated $N$ vortex cores in each condensates : 
\Align{Initial_Guess1}{
\psi_1&= |\psi_1|\mathrm{e}^{ i\Theta} \, ,
\psi_2= |\psi_2|\mathrm{e}^{ i\Theta+i\Delta_{12}} \, ,
\psi_3= |\psi_3|\mathrm{e}^{ i\Theta+i\Delta_{13}} \, ,~~  \nonumber \\
|\psi_a| &= u_a\prod_{i=1}^{N_v} 
\sqrt{\frac{1}{2} \left( 1+\tanh\left(\frac{4}{\xi_a}({\cal R}_i(x,y)-\xi_a) \right)\right)}\, ,~~  \nonumber \\
\mbf{A}&=
\frac{1}{e{\cal R}}\left(\sin\Theta,-\cos\Theta \right)\,,
}
where $a=1,2,3\,$ and $u_a\,$ is the ground state value of each superfluid density. The parameter $\xi_a$ gives 
the core size while $\Theta\,$ and $\cal{R}\,$ are 
\Align{Initial_Guess2}{
\Theta(x,y)&=\sum_{i=1}^{N_v}\Theta_i(x,y) \,,  \nonumber\\
\Theta_i(x,y)&=\tan^{-1}\left(\frac{y-y_i}{x-x_i}\right)\,, \nonumber\\
{\cal R}(x,y)&=\sum_{i=1}^{N_v}{\cal R}_i(x,y)\,, \nonumber\\
{\cal R}_i(x,y)&=\sqrt{(x-x_i)^2+(y-y_i)^2}\,. 
}
The initial position of a  vortex is given by $(x_i,y_i)$. The  functions $\Delta_{ab}\equiv\varphi_b-\varphi_a$ can be 
used to initiate a domain wall. As an initial guess we generally choose $\Delta_{12}=-\Delta_{13}\equiv\Delta$, with 
$\Delta$ defined as
\Equation{PhaseDiff}{
   \Delta=\frac{\pi}{3}\left( H({\mbf r}-{\mbf r}_0)-1\right)\,,
}
where  $H({\mbf r}-{\mbf r}_0)$ is a Heaviside function.  Thus in the initial guess  the domain wall has infinitesimal 
thickness. It takes only a few steps from this initial guess to relax to a true domain wall during the simulations. 
Consequently, it is entirely sufficient to use Heaviside functions for the initial guesses of domain walls. Once the initial 
configuration defined, all degrees of freedom are relaxed simultaneously, within the `open' boundary conditions 
discussed previously, to obtain highly accurate solutions of the Ginzburg-Landau equations. In a strongly type-II 
system when the initial guess was  either (a) vortices placed on a domain wall or (b) closed domain wall surrounding 
a densely packed group of vortices, the system almost always formed $\CPtwo$  solitons. We used also initial guesses 
(c) without any domain walls ($\Delta=0$). In that case we observed $\CPtwo$ soliton formation, if in the initial states 
vortices were densely packed. This again indicating that the $\CPtwo$ solitons in the three component GL model 
represent local minima with wide capture basin in the free energy landscape.

 Figure 7 in the paper shows stages of the energy minimization. Corresponding movies are available as the supplementary 
 online material \cite{Julien}. The main focus of this work is the existence of stable static solutions, however the numerical 
relaxation scheme which we use can give insight into possible formation dynamics of these objects. That is, 
the gauge invariant gradient flow of Ginzburg-Landau free energy can be related to the dynamics of Time Dependent 
Ginzburg-Landau equations  \cite{Note2,Du:94}. Therefore supplementary movies not only give  
information about the size of the capture basin of the local minima associated with the $\CPtwo$  solitons, they also 
provide some insight into possible real dynamics which can lead to their formation.

\section{Appendix B : Stability of the solutions}\label{Numerics-stab}

\begin{figure*}[!htb]
  \hbox to \linewidth{ \hss
  \includegraphics[width=\linewidth]{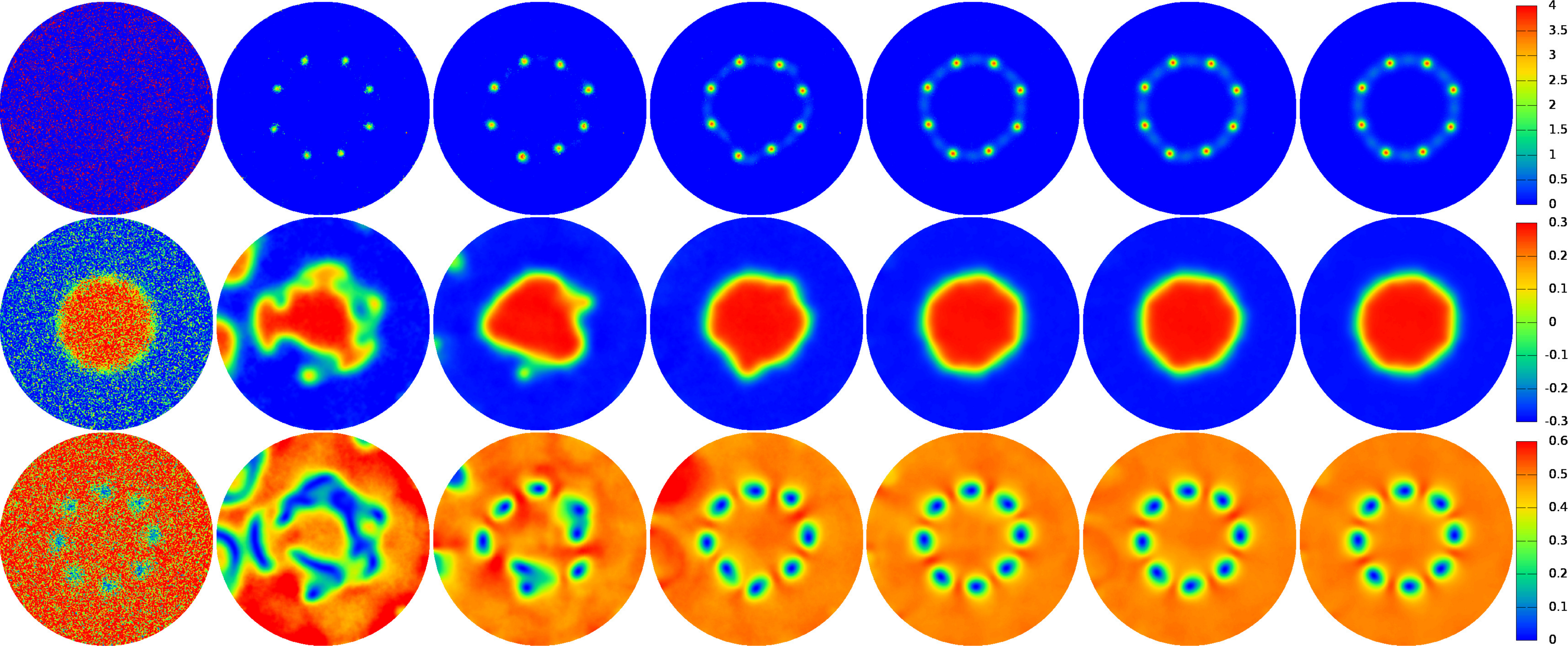}
  \hss}
\caption{
(Color online) -- Displayed quantities are the same as in Fig.~7 of the paper, namely the 
energy density, $(\psi_1^*\psi_2-\psi_1\psi_2^*)/2i$ 
and $|\psi_1|^2$. Initial configuration is a charge $8$ $\CPtwo$ soliton shown in Fig.~3. 
The snapshots show the state of the system a different stages of the energy minimization 
algorithm after the applied perturbation. The initial noise is $P=0.6$, which in fact is a very  significant perturbation
where the density fields vary locally  up to 60 \% of the ground state values, while
magnetic field varies up to 60 \% of its maximal value. 
The configuration nevertheless relaxes back to the $\CPtwo$ soliton.
}
\label{Stability-fig1}
\end{figure*}

\begin{figure*}[!htb]
  \hbox to \linewidth{ \hss
  \includegraphics[width=\linewidth]{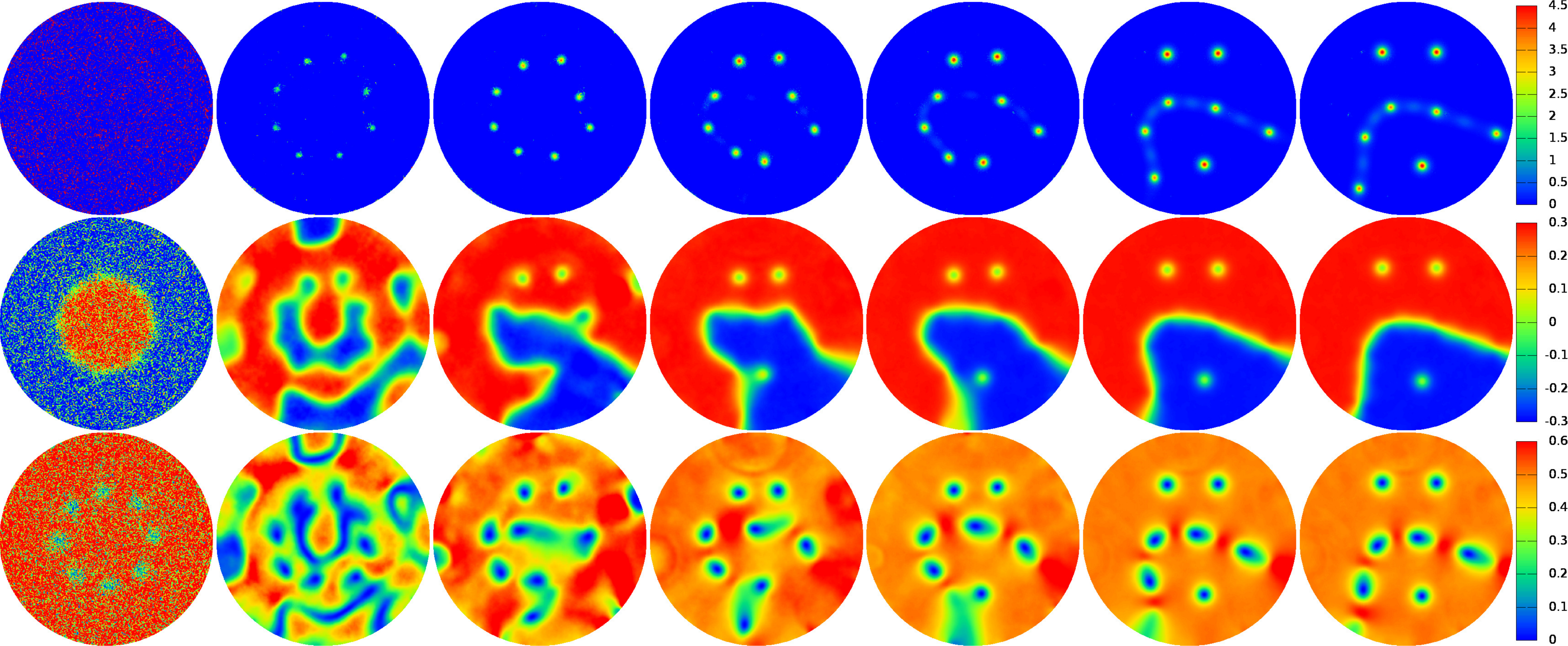}
  \hss}
\caption{
(Color online) -- Displayed quantities as well as the initial solution are the same as in \Figref{Stability-fig1}. Now the initial 
noise is $P=0.7$. Here the noise is strong enough to open a hole in the domain wall, which then emits ordinary vortices
 and decays by being absorbed by the 
boundary of the domain.  
}
\label{Stability-fig2}
\end{figure*}

The solutions were obtained using an (energy) minimization algorithm, and  {not} by solving the equations of 
motion. As a result, after the convergence (which is carefully controlled), the solution is guaranteed to represent 
(at least) a \emph{local} minimum of the energy functional \Eqref{freeEnergy}. Because no symmetry-imposing 
ansatz is used,  there are no possible unstable modes   truncated by symmetry assumptions.  
Linear stability analysis consists of applying  infinitesimally small perturbation to the fields, and investigating 
the eigenvalue spectrum of the (linear) perturbation operator, on the background of a given soliton. When the 
background solution is (meta) stable all infinitesimally small perturbations are positive modes and thus can only 
increase the energy. However a strong perturbation should cause a decay of a soliton to ordinary vortices
since these solitons are protected against decay by a finite energy barrier. Instead of studying different modes, 
we double-checked the stability numerically  by perturbing the solution by a random noise. The random noise which 
is applied to all degrees of freedom, is generated as follows 
\Align{Perturbations}{
   \Re(\psi_a) &= \Re(\psi_a)^\oz + Pu_a\mu_a^{\mbox{\tiny Re}}(x,y)\, ,   \nonumber \\
   \Im(\psi_a) &= \Im(\psi_a)^\oz + Pu_a\mu_a^{\mbox{\tiny Im}}(x,y)\, ,  \nonumber \\
   A_i&= A_i^\oz+P\mathrm{max}(|\mbf A|)\mu_i^{\mbox{\tiny A}}(x,y)	\,.
}
Here $^\oz$ denotes the background solutions, $P$ is a percentage giving the relative magnitude of the fluctuation with 
respect to the maximal amplitude of a given field of the background solution. $\mu_a^{\mbox{\tiny Re}}(x,y)$, 
$\mu_a^{\mbox{\tiny Im}}(x,y)$ and $\mu_i^{\mbox{\tiny A}}(x,y)$ are (independent) random functions of the 
space $\in[-1:1]$. As a result all fields initially receive  noise whose relative amplitude is $P$. The system is then again 
relaxed using the same minimization scheme as for constructing the solitons. It is found that if  the random noise does 
not exceed a certain threshold, the configuration relaxes back to the soliton solution, as can be seen from \Figref{Stability-fig1}.
The noise was gradually increased, finding that indeed, sufficiently strong perturbation drives the soliton over the barrier, in 
the energy landscape. Thus leading to its decay to ordinary vortex solutions as shown on \Figref{Stability-fig2}. The precise 
value of the relative amplitude required to destabilize a given soliton, obviously depend  on the GL parameters and on the number 
of flux quanta of the solution.

%%%%%%%%%%%%%%%%%%%%%%%%%%%%%%%%%%%%%%%%%%%%%%%%%%%%%%%%%%%%%%%%%%%%%%%%%%%%%%%%%%%%%%%%%%
%%%% Bibliography
%Merlin.mbs v4.21 2009-07-09.
%


\begin{thebibliography}{10}
\bibitem{remark}%
  \BibitemOpen
  \bibinfo {note} {
For this study, we investigated about $600$ different configurations, with an average of $100$ CPU 
 hours per invested configuration. Which required computational resources of a very 
 large supercomputer.}%
  \bibAnnoteFile{Stop}{remark}%
\end{thebibliography}
\end{document}